\documentclass[a4paper,fleqn,usenatbib]{mnras}

\usepackage{amsmath}  % Advanced maths commands
\usepackage{amssymb}  % Extra maths symbols

\usepackage{mathptmx} 
\usepackage{txfonts}

\usepackage[T1]{fontenc}
\usepackage{ae,aecompl}

\usepackage{graphicx}	% Including figure files

\usepackage{listings}

\title[Uncertain Photometric Redshifts]{Uncertain Photometric Redshifts}

\author[Polsterer, D'Isanto, and Gieseke]{
Kai Lars Polsterer,$^{1}$\thanks{E-mail: kai.polsterer@h-its.org (KLP)}
Antonio D'Isanto,$^{1}$
and Fabian Gieseke,$^{2}$
\\
$^{1}$Heidelberg Institute for Theoretical Studies (HITS) gGmbH, Astroinformatics, Schloss-Wolfsbrunnenweg 35, 69118 Heidelberg, Germany\\
$^{2}$Radboud University, Institute for Computing and Information Sciences, Heyendaalseweg 135, 6525 AJ Nijmegen, the Netherlands\\
}

\date{Accepted XXX. Received YYY; in original form ZZZ}

\pubyear{2016}

\begin{document}
\label{firstpage}
\pagerange{\pageref{firstpage}--\pageref{lastpage}}
\maketitle

% Abstract of the paper
\begin{abstract}
Photometric redshifts play an important role as a measure of distance for various cosmological topics.
Spectroscopic redshifts are only available for a very limited number of objects but can be used for creating statistical models.
A broad variety of photometric catalogues provide uncertain low resolution spectral information for galaxies and quasars that can be used to infer a redshift.
Many different techniques have been developed to produce those redshift estimates with increasing precision.
Instead of providing a point estimate only, astronomers start to generate probabilistic density functions (\emph{PDF}s) which should provide a characterisation of the uncertainties of the estimation.
In this work we present two simple approaches on how to generate those \emph{PDF}s.
We use the example of generating the photometric redshift \emph{PDF}s of quasars from \emph{SDSS}(\emph{DR7}) to validate our approaches and to compare them with point estimates.
We do not aim for presenting a new best performing method, but we choose an intuitive approach that is based on well known machine learning algorithms.
Furthermore we introduce proper tools for evaluating the performance of \emph{PDF}s in the context of astronomy.
The continuous ranked probability score (\emph{CRPS}) and the probability integral transform (\emph{PIT}) are well accepted in the weather forecasting community.
Both tools reflect how well the \emph{PDF}s reproduce the real values of the analysed objects.
As we show, nearly all currently used measures in astronomy show severe weaknesses when used to evaluate \emph{PDF}s.

\end{abstract}

% Select between one and six entries from the list of approved keywords.
% Don't make up new ones.

%%%%%%%%%%%%%%%%%%%%%%%%%%%%%%%%%%%%%%%%%%%%%%%%%%

%%%%%%%%%%%%%%%%% BODY OF PAPER %%%%%%%%%%%%%%%%%%

\section{Introduction}

Determining the distance of an object has always been an important task in astronomy.
This involves distances on all scales, from our solar system to distances on cosmological scale.
Generating measures with high confidence requires dedicated instrumentation and an appropriate observation time.
Unfortunately, obtaining precise spectroscopic distance measures for all objects is rendered impossible by both, the huge amount of required observation time and the necessary sensitivity of the instrumentation.
Especially for cosmology, precise distances are required to understand e.g. gravitational lensing on cosmic scales \citep{2005ApJ...624...59H,2005ApJ...633..589S}.
For example, the success of the \emph{Euclid} mission \citep{2010SPIE.7731E..1HL} will heavily depend on the availability of precise distance measures.
The distance on cosmological scale, the so called redshift $z$ that is caused by the expansion of the universe, has therefore be determined through alternative methods.

Based on the huge amount of data that is made available by dedicated survey mission, statistical methods can be used to infer redshifts.
Typically, a wide range of photometric measurements are extracted from the imaging surveys like \emph{SDSS} \citep{2000AJ....120.1579Y}, \emph{UKIDSS} \citep{2007MNRAS.379.1599L}, and \emph{WISE} \citep{2010AJ....140.1868W}.
Due to the usually selected broadband filters, multiple measurements allow to very broadly reconstruct the spectral energy distribution (\emph{SED}).
In the past, \emph{SED} fitting approaches \citep[as][]{2000A&A...363..476B} have provided very good estimations for the redshift based on photometric measurements.
\citet{2003AJ....125..580C} utilized a variety of techniques to generate photometric redshift estimates for the early data release of \emph{SDSS}, while \citet{2016MNRAS.460.1371B} provide redshifts for the latest release.
Besides classical template fitting approaches, more and more machine learning approaches are used.
Instead of using a fixed set of template spectra those approaches are based on models trained on a given reference set \citep{2011MNRAS.418.2165L,2013MNRAS.428..226P,2014A&A...568A.126B}.
There is a wide range of algorithms available including artificial neural networks, random forest, nearest neighbour approaches or support vector regression models.
In \citet{bishop2007} and \citet{hastie_elements_2001}, introductions to various machine learning techniques are provided. 

Instead of just generating a point estimate for a photometric redshift, some approaches can generate a probability density function (\emph{PDF}).
In contrast to dealing with just a single prediction, this enables scientists to evaluate the likelihood for different redshifts.
By integrating over a certain region of the \emph{PDF}, the probability of being in a certain redshift range can be directly calculated.
This measure of likelihood improves the analysis and further usage of the predictions, as the uncertainty can be quantified and propagated correctly.
Important for a \emph{PDF} is that the integral is always one, otherwise it would not be a correct density distribution.

The generation and evaluation of \emph{PDF}s introduces new challenges.
In this work we introduce proper tools to astronomy, that allow a fair comparison of the prediction quality achieved through \emph{PDF}s.
By discussing common errors, we want to emphasize the necessity of proper tools and measures.
In addition, we present two simple approaches for generating \emph{PDF}s based on well known algorithms, a nearest neighbours approach and a random forest regression model.
We do not claim, that these methods produce the best results, they are just simple examples for generating \emph{PDF}s.
Feature selection and post-processing are appropriate tools to further improve the generated \emph{PDF}s \citep{2014ASPC..485..425P}.

\vspace{1em}
\noindent
{\bfseries Outline:}
After this brief introduction to the topic of photometric redshift estimation, in \mbox{Section \ref{quality}} we will discuss proper tools for estimating the quality of \emph{PDF}s.
Next we present two straightforward approaches for creating \emph{PDF}s based on well known algorithms (see \mbox{Section \ref{generating}}).
In \mbox{Section \ref{experiments}} the presented approaches and evaluation tools are used to demonstrate their performance in some simple redshift estimation experiments.
We also discuss the results of the experiments in detail.
Finally, in \mbox{Section \ref{conclusion}}, we conclude our work.

\section{Quality Estimation Tools}
\label{quality}

When it comes to evaluating \emph{PDF}s, one has to rethink the tools used for measuring the quality of an estimation.
For the task of photometric redshift estimation, we make use of reference data taken via spectroscopy.
Therefore we have training, test and evaluation data-sets that map photometric features to highly certain and unique redshifts.
This allows us to measure the performance of our redshift estimation models very precisely.
To be able to compare our results we should use a common set of measures and tools.

One of the currently proposed extensions for generating \emph{PDF}s, is to simply add an uncertainty interval to a generated point estimate.
Assuming a normal distribution, one could use this formulation to calculate a certain likelihood for every possible redshift.
The task of evaluating this distribution with respect to a single photometric redshift cannot be done by using the measures that have been used in the past.
We propose a new set of measures and tools that are capable of evaluating \emph{PDF}s with respect to a single true value.
Before giving a detailed description we would like to point out some common mistakes with respect to the \emph{PDF} evaluation process.

\subsection{Common Mistakes}
\label{mistakes}

In the past a lot of measures have been used to quantify the characteristics of the error distribution.
It was easy to measure the deviation between the true value and an estimated value.
Normalization with respect to the redshift was commonly applied to express a relative error.
The mean/median of the error distribution was a very simple statistical measure to check for biases, while other measures were used to quantify the spread across the deviations.
Those measures are well suited for analysing and describing the deviations between point estimates and true values.

\subsubsection{Oversimplifying \emph{PDF}s}

\begin{figure}
  \includegraphics[width=\columnwidth]{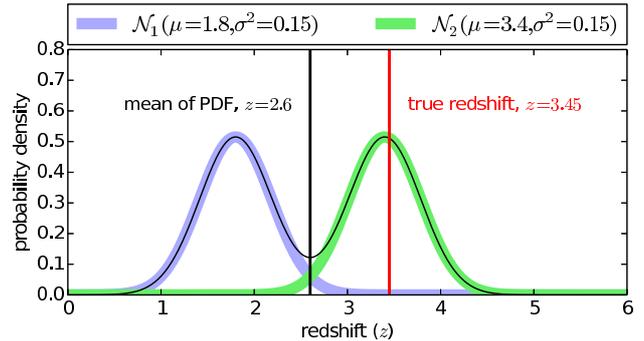}
    \caption{Example of a wrong evaluation of a \emph{PDF} via its mean.
    A bimodal \emph{PDF} which is composed out of two normal distributions $\mathcal{N}_1$ and $\mathcal{N}_2$ is simplified to its mean value.
    When comparing the mean value $z=2.6$ to the true redshift $z=3.45$, the difference becomes obvious, as the true redshift is much closer to the centre of $\mathcal{N}_2$ than it is to the mean.
    }
    \label{fig:mistakes1}
\end{figure}

Most of these measures have been commonly applied and are well understood and accepted by the community.
Therefore it can be understood that they are applied in modified versions to evaluate \emph{PDF}s.
In order to apply those well known measures, e.g. the mean of a \emph{PDF} is extracted and used as a single value for comparison.
A simple example of why this approach is not appropriate is shown in \mbox{Figure \ref{fig:mistakes1}}.
Multimodal distributions are common for photometric redshift estimation problems, due to the low photometric wavelength resolution, differences in the nature of the underlying physical sources, and measurement uncertainties.
In \mbox{Section \ref{generating}}, we present a straightforward approach to generate \emph{PDF}s.
This approach clearly demonstrates that we have to expect multimodal distributions.
In case of a simple bimodal distribution it is clear that the mean is likely to be in a region in between of the two components and therefore will exhibit a relatively low probability density.
Calculating the root mean square error between the mean of the \emph{PDF} and the true value is hence producing useless information.
A measure based on the mean would misleadingly indicate a good performance, as soon as a non symmetric \emph{PDF} is evaluated.
To prevent confusion, especially if the true nature of the produced \emph{PDF}s is unclear, {\bf we do not recommend the use of this measure}, especially when comparing the results of different approaches.
As the \emph{PDF}s are defining a probability distribution, this information must be used accordingly and not over-simplified because of a preferred measure.

\subsubsection{Wrong Definition of Outliers}

\begin{figure*}
  \includegraphics[width=\textwidth]{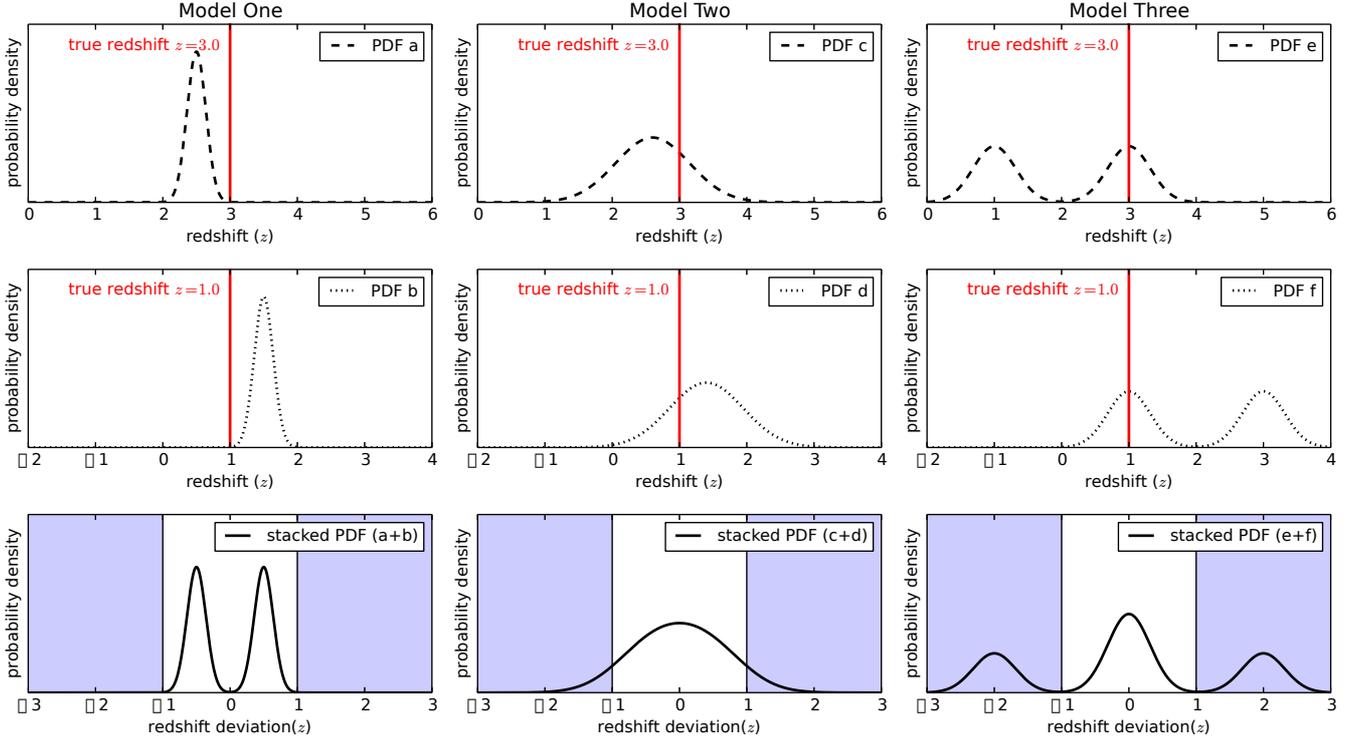}
    \caption{Example for a wrong definition of outliers.
    In Model One (left), two narrow single Gaussian \emph{PDF}s (a,b) for two different objects are co-added.
    Both \emph{PDF}s were shifted by the true redshift, to be aligned before stacking.
    In the resulting stacked \emph{PDF}, values outside $[-1,1]$ are considered to be outliers with respect to their estimation error.
    Even though the true redshift value is not well represented by the stacked \emph{PDF}, the plot shows no outliers.
    In Model Two (middle), two broader single Gaussian \emph{PDF}s (c,d) are stacked.
    Even though both \emph{PDF}s represent very well the true redshift, the stacked version exhibits a certain fraction in the marked outlier region.
    Model Three (right) shows the result of stacking two bimodal \emph{PDF}s (e,f) with the true redshift always being perfectly matched by one of the peaks.
    As it is shown in the resulting plot 50 percent of the surface of the stacked plot is out of the desired range.
    This demonstrates an obvious shortcoming of this method.
    }
    \label{fig:mistakes2}
\end{figure*}

Catastrophic outliers with respect to the error are considered to be bad when it comes to estimating photometric redshifts.
In the past, an outlier was defined to be a point estimate that is too far away from its true redshift.
The distribution of the deviations between the estimated and the true redshift value allowed a simple characterisation of those outliers by defining simple thresholds.
Typically a histogram was used to inspect the distribution of the deviations.
In \citet{2011arXiv1110.3193L}, the requirements of the \emph{Euclid} mission are collected.
The normalised standard deviation of the photometric redshift estimation $\sigma_z / (1 + z)$ is specified to be lower than 5 percent for not more than 10 percent of the objects.
In this section, we do not want to discuss whether this target is reachable or not.
We would like to highlight the problems that arise by sticking to such a measure of outliers, when using \emph{PDF}s instead of point estimates.
In \mbox{Figure \ref{fig:mistakes2}} we visualise the problems with such a measure.
To be able to measure the amount of outliers based on a threshold criterion, we observe that typically \emph{PDF}s are shifted by their true redshift and co-added.
The stacked result of all evaluated \emph{PDF}s allows to determine the amount of predictions that are exceeding a given range.
This can be seen to be similar to the evaluation of the histograms created for the deviations of the point estimates.
In our example we choose to have three different models that generate \emph{PDF}s for two objects, respectively.
Model One generates two narrow Gaussian shaped \emph{PDF}s that are shifted against the true redshift.
Even though the resulting stacked \emph{PDF} shows no outliers, it is obvious that the result is terribly off, as the likelihood for the true redshift is practically zero.
As soon as broader Gaussians are used, the true redshift is better represented as indicated by the higher likelihood at the true redshift.
Even though both \emph{PDF}s of Model Two are more likely, the stacked \emph{PDF} shows already a certain area in the marked outlier region.
In our last example we choose to stack two bimodal \emph{PDF}s.
As one can see from Model Three, the resulting stacked \emph{PDF} has more than 50 percent of its area outside the given range.
Even though, each \emph{PDF} has a very high density at the true redshift, the stacked version would indicate a catastrophicly high rate of outliers.

When dealing with \emph{PDF}s, {\bf outliers should be defined based on probability and not on the area of the stacked \emph{PDF}}.
Even though the similarity is very high to the method commonly applied to point estimates, it cannot be used to quantify outliers based on \emph{PDF}s.
A \emph{PDF} that is excluding the true redshift by showing a too low likelihood at this point, must be considered a catastrophic outlier instead.
Not the relative difference between the true redshift and parts of the \emph{PDF}, but the likelihoods should be used as an outlier criterion. 

\subsection{Likelihood}
\label{likelihood}

Given a \emph{PDF} as a redshift estimation and the true redshift value $z$, it is possible to evaluate the likelihood of the \emph{PDF} at the given value.
This likelihood specifies how well the given \emph{PDF} represents the true redshift value.
A low likelihood clearly indicates a unlikely redshift.
In case the likelihood of an estimated \emph{PDF} at the true redshift is very low, the \emph{PDF} is not a good representation.
Therefore the likelihood can be used to express how well a \emph{PDF} represents the true redshift value.
By maximizing the likelihood of the \emph{PDF}, we can improve our model.

Based on the likelihood, we can efficiently detect those redshifts where the \emph{PDF} fails to represent the true redshift.
A histogram of the likelihoods of the \emph{PDF}s at the corresponding true redshift value $z$, visualizes this distribution and allows to define a cutoff criterion.
Every \emph{PDF} below such a threshold, can be considered a catastrophic outlier with respect to the photometric estimation.

\subsection{Probability Integral Transform}
\label{PIT}

\begin{figure*}
  \includegraphics[width=\textwidth]{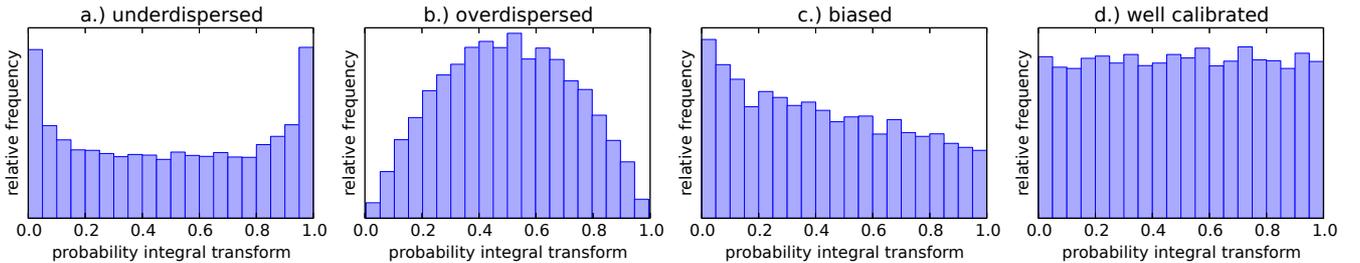}
    \caption{Four different probability integral transforms (\emph{PIT}s).
    In the case of underdispersed \emph{PDF}s an u-shaped, concave distribution is observed (a).
    Overdispersed \emph{PDF}s result in a peaked, convex distribution (b).
    When a slope in the \emph{PIT} is observed, the analysed \emph{PDF}s are biased (c).
    Only when the \emph{PIT} exhibits a flat distribution, the \emph{PDF}s are well calibrated (d).
    }
    \label{fig:pit}
\end{figure*}

As stated by \citet{gneiting2007probabilistic}, when comparing forecasting distributions and observations, the goal is to maximize the sharpness of the predictive distributions subject to calibration.
In the context of photometric redshift estimation this refers to comparing \emph{PDF}s with spectroscopic redshifts.
The term calibration describes the consistency between the predictive distribution and the true redshift.
Sharpness is used to express the concentration of the \emph{PDF}.

In \citet{dawid1984present} the probability integral transform (\emph{PIT}) was proposed to be used as a diagnostic tool to check the calibration and the sharpness of the generated predictive distributions.
The \emph{PIT} is a visual tool which is based on the histogram of the values of the cumulative probability at the true value.
Therefore the \emph{PDF}s have to be transferred into continuous density functions (\emph{CDF})s \mbox{(see Equation \ref{eqn:CDF})}.

\begin{equation}
CDF_{t}(z_{t}) = \int_{- \infty}^{z_{t}}PDF_{t}(z)dz
\label{eqn:CDF}
\end{equation}

\noindent
With respect to photometric redshift estimations, the \emph{PIT} is calculated with the \emph{CDF} of the estimated redshift $CDF_t$ at the true redshift $z_{t}$ (see \mbox{Equation \ref{PIT}}).
Hereby $t\in \{1,2,\ldots N\}$ indexes the corresponding tuple of a predicted redshift distribution and the matching true redshift for $N$ data items.

\begin{equation}
p_{t} = CDF_{t}(z_{t})
\label{PIT}
\end{equation}

\noindent
In \mbox{Appendix \ref{pitExampleCode}} an example is provided on how to generate a \emph{PIT} histogram based on plain predictions.
When the anaylsed \emph{PDF}s are of Gaussian nature with $\mu$ and $\sigma^2$ as mean and variance, the \emph{CDF}s can be evaluated by using \mbox{Equation \ref{CDF}}.
For a Gaussian mixture model, the corresponding \emph{CDF} is a additive mixture of single Gaussian \emph{CDF}s multiplied with their weights, respectively.

\begin{equation}
CDF_{t}(z_{t}) = \frac{1}{2}\left[1+erf\left(\frac{z_{t}-\mu}{\sqrt{2\sigma^2}}\right)\right]
\label{CDF}
\end{equation}

\noindent
In case the predictions are ideal, the distribution of $p_{t}, t \in \{1,2,\ldots N\}$ has to be uniform.
As shown in \mbox{Figure \ref{fig:pit}}, multiple aspects can be verified by plotting the histogram of this distribution.
Only if the distribution of $p_{t}$ exhibits a uniform shape, the \emph{PDF}s are well calibrated.
When the dispersion of the estimates is to small in relation to the distribution of the true redshifts, an underdispersed distribution of $p_{t}$ can be observed.
This will be reflected by a u-shaped, concave histogram.
The opposite case is observed with overdispersed \emph{PDF}s that generate a peaked, concave histogram.
As soon as a bias is present in the \emph{PDF}s, a slope is added to the distribution of $p_{t}$.

The histogram of the \emph{PIT} values allows to visually check the calibration and sharpness.
It provides intuitive access to multiple aspects of the \emph{PDF}s with respect to the corresponding true redshifts.
Therefore we recommend this tool to be always used when evaluating \emph{PDF}s.

\subsection{Continuous Ranked Probability Score}
\label{CRPS}

\begin{figure*}
  \includegraphics[width=\textwidth]{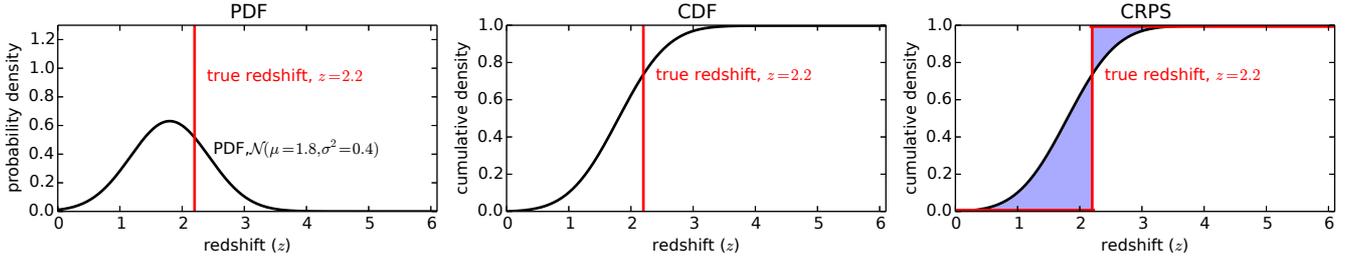}
    \caption{Visual guide to the meaning of probability density function (\emph{PDF}), continuous density function (\emph{CDF}), and continuous ranked probability score (\emph{CRPS}).
    The real redshift that is used for calculating the \emph{CRPS}, is plotted as a reference in red.
    The integral over and under the (\emph{CDF}) that is used for the calculation of the \emph{CRPS} is coloured in blue.
    Note the Heaviside step-function of the true redshift that is marked red in the \emph{CRPS} plot (right).
    }
    \label{fig:crps}
\end{figure*}

When comparing the performances of different approaches that generated \emph{PDF}s based on photometric features, a proper score should be used to measure the individual performances.
Please see \citet{RePEc:bes:jnlasa:v:102:y:2007:p:359-378} for a detailed introduction to the topic of proper scoring rules.
In this work we make use of the continuous ranked probability score (\emph{CRPS}) as a performance measure.
The \emph{CRPS} \citep{Hersbach} is widely used in the field of weather forecasting for expressing a distance between a \emph{PDF} and a true value.
It compares a full distribution with an observation as defined in :

\begin{equation}\begin{split}
CRPS = \frac{1}{N}\sum_{t=1}^{N}crps(CDF_{t}, z_{t}), \\
\text{with } crps(CDF_{t}, z_{t}) = \int_{- \infty}^{+ \infty}\left[CDF_{t}(z) - CDF_{z_{t}}(z)\right]^{2}dz
\label{eqn:CRPS}
\end{split}
\end{equation}

\noindent
$CDF_{t}$ is the cumulative distribution of the \emph{PDF}, as defined in \mbox{Equation \ref{eqn:CDF}}.
In \mbox {Equation \ref{eqn:CDFz}} the cumulative distribution of the true redshift $CDF_{z_{t}}$ is defined based on $H(z) = \mathcal{H}$, the Heaviside step-function.

\begin{equation}
CDF_{z_{t}}(z) = H(z - z_{t}), \text{ with } H(z) = \left\{
    \begin{array}{lr}
      0 & \text{for } z < 0 \\
      1 & \text{for } z \geq 0
    \end{array}
    \right. 
\label{eqn:CDFz}
\end{equation}

\noindent
The calculation of the \emph{CRPS} as well as a Gaussian \emph{PDF} and the corresponding \emph{CDF} are visualised in \mbox{Figure \ref{fig:crps}}.
In case the \emph{PDF}s are given as normal distributions, we are able to write it in the subsequent form \citep{Gneiting2005}.

\begin{equation}\begin{split}
crps[\mathcal{N}(\mu_{t}, \sigma_{t}^{2}), z_{t}] =\\
\sigma_{t}
\left \{
\frac{z_{t} - \mu_{t}}{\sigma_{t}} \left[2\Phi\left(\frac{z_{t} - \mu_{t}}{\sigma_{t}}\right) - 1\right]
 + 2\phi\left(\frac{z_{t} - \mu_{t}}{\sigma_{t}}\right) - \frac{1}{\sqrt{\pi}}
 \right \},
\end{split}
\label{eqn:CRPScalc}
\end{equation}

\noindent
where $\phi$ and $\Phi$ represent the \emph{PDF} and the \emph{CDF} of a normal distribution with mean $0$ and variance $1$, respectively.
In \mbox{Equation \ref{eqn:CRPScalc}} the $\frac{z_{t} - \mu_{t}}{\sigma_{t}}$ term represents the normalized prediction error.
Representing a \emph{PDF} as a Gaussian mixture model (\emph{GMM}) \citep{bishop2007} provides some advantages in calculating the \emph{CRPS} for even more complicated distribution.
A Gaussian mixture model (see \mbox{Equation \ref{eqn:GMM}}) defines a distribution as a combination of $M$ number of Gaussians with independent means ${\bf \mu}$ and variances ${\bf \sigma}^2$.
\begin{equation}\begin{split}
GMM({\bf \mu}, {\bf \sigma}^2, {\bf \omega}) = \sum_{i=1}^{M} \omega_{i} \mathcal{N}(\mu_{i},\sigma_{i}^{2}),\\
\text{ with } \sum_{i=1}^{M}\omega_{i}=1 \text{ and } \omega_{i} \geq 0, \forall i \in \{1,2,\ldots M\}
\end{split}
\label{eqn:GMM}
\end{equation}

\noindent
hereby the weights ${\bf \omega}$ control the contributions of the individual Gaussians to the final distribution.
With

\begin{equation}
A(\mu,\sigma^2) = 2 \sigma \phi \left ({\frac{\mu}{\sigma}} \right ) + \mu \left [ 2 \Phi\left(\frac{\mu}{\sigma} \right )   -1 \right ] \text{, and}
\label{eqn:A}
\end{equation}

\begin{equation}
\begin{split}
crps(GMM_t({\bf \mu}, {\bf \sigma}^2, {\bf \omega}), z_t) =\\ 
\sum_{i=1}^{M} \omega_{i} * A(z_{t}- \mu_{i},\sigma_{i}^2) - \sum_{i=1}^{M} \sum_{j=1}^{M} \frac{1}{2} \omega_{i}\omega_{j} * A(\mu_{i}-\mu_{j},\sigma_{i}^2-\sigma_{j}^2)
\label{eqn:CRPSgmm}
\end{split}
\end{equation}

\noindent
we can calculate the \emph{CRPS} of a \emph{GMM} \citep{QJ:QJ2006132621104}.
For a lot of scores the {\tt properscoring} package provides a \emph{Python} implementation that can be easily used for calculating the \emph{CRPS}.
This package includes the calculation of the \emph{CRPS} for an ensemble of predictions, too.

\section{Generating Uncertain Redshifts}
\label{generating}

In the past, multiple approaches have been developed and applied to whole zoo of photometric features to generate redshift point estimates.
Some of those approaches have been extended to generate \emph{PDF}s by either sampling over the input uncertainties or adding fixed uncertainties to the point estimates.
When fitting spectral energy distributions to the photometric features, under certain limitations, \emph{PDF}s can be derived directly.
Instead of comparing or discussing those approaches, we would like to introduce two simple but powerful approaches to generate \emph{PDF}s.

\subsection{Nearest Neighbours based \emph{PDF}}

\begin{figure}
  \includegraphics[width=\columnwidth]{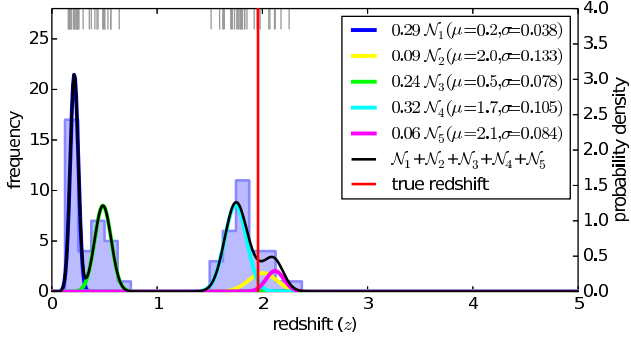}
    \caption{\emph{PDF} of a quasar at $z=1.959$ calculated with a reference set of $30,000$ high redshift quasars taken from \emph{SDSS}(\emph{DR7}).
    Based on the 64 nearest neighbours only, a histogram of the redshift distribution is generated (background).
    The individual redshifts of the neighbours are marked at the upper edge.
    The fitted Gaussian mixture model and its components are plotted in the foreground.
    Note the obvious multimodal distribution of the neighbours that reproduces the true redshift (marked in red) quite well.
    }
    \label{fig:nn}
\end{figure}

To generate a point estimate, nearest neighbour based approaches have proven to generate good results.
Based on the redshift of the nearest neighbours a point estimate is generated, usually by either calculating the mean or the weighted mean.
This approach performs very well for a large set of reference objects, as a larger set is more likely to nicely represent all the objects in the high-dimensional feature space.
Because this approach can be seen as sampling the feature space in the vicinity of an object, we assume that this sampling can be used to estimate the density of the distribution with respect to the redshift $z$.
Instead of calculating a single point estimate, we use the neighbours as a density representation.

By fitting a Gaussian mixture model (\emph{GMM}) we can transfer the individual redshift values of the neighbours into a Gaussian representation.
Therefore we can use the nearest neighbours to produce a \emph{PDF} instead of a point estimate.
See \mbox{Figure \ref{fig:nn}} for an example of fitting a \emph{GMM} to the nearest neighbours redshift distribution in order to generate a \emph{PDF}.
This example is based on the task of fitting photometric redshifts of quasars taken form \emph{SDSS}(\emph{DR7}).
When checking the distribution of the nearest neighbours, a multimodal distribution can be observed.
In our example, a point estimate would be in the redshift region around $z=1$ even though neither the true redshift nor the neighbours are close to $z=1$.
This is a clear indicator, why a plain point estimate will never be able to produce a good representation of the true redshift.
An example on how to formalise this in \emph{Python} is given in \mbox{Appendix \ref{exampleCode}}.

It is obvious that this approach suffers from one limitation.
In order to fit the \emph{GMM} quite well, we need a large number of neighbours that are not too far away with respect to the applied distance measure in the feature space.
\mbox{Section \ref{experiments}} provides an analysis of the number of considered neighbours with respect to the number of fitted Gaussians.
Especially those regions in the feature space that show a low number of reference objects because of physical or observational selection effects, will produce bad redshift estimations.
One could consider weighting the individual neighbours when fitting the \emph{GMM}, but still the number of neighbours that represent the redshift well, will be low.
This will either smear out the resulting \emph{PDF} by covering a too large range of references or produce inaccurate \emph{PDF}s by over-fitting the \emph{GMM}.
As this approach is very intuitive and helpful for comparing the performance of other approaches, we use it to retrieve a base performance measure.
The approach presented next is not so sensitive to the discussed limitation.

\subsection{Random Forest based \emph{PDF}}

A random forest is an ensemble of uncorrelated decision trees that are created based on a given set of reference objects.
More details on random forest regression models can be found in \citet{Breiman:2001:RF:570181.570182}.
Typically this ensemble is used to derive a mean of all members.
Each decision tree is partitioning the feature space.
Similarly to the nearest neighbour approach, the individual member values can be interpreted as a part of a redshift distribution.
Therefore we fit a \emph{GMM} directly to the results of the individual decision trees, instead of deriving just a mean value.
In \mbox{Appendix \ref{exampleCode}} we describe how to modify the nearest neighbour example to use a random forest regressor.

In contrast to the nearest neighbour approach, we are not strictly bound to the number of nearest neighbours to sample our high-dimensional feature space.
The randomisation during the creation of the decision trees via feature selection, bootstrapping, and bagging creates independent trees which partition the feature space.
By producing a larger number of trees, we are not smearing out the redshift information as we did for the nearest neighbour approach.
We do not sample directly from our reference set, we use independent representations instead.

In the experiments section this difference in sampling from the feature space is directly reflected by the performance of the resulting \emph{PDF}s.

\section{Experiments and Results}
\label{experiments}

In order to evaluate the performance of the introduced approaches, we will focus on the use-case of photometric redshift estimation of quasars.
To provide results that are easily comparable to other publications we choose to use the photometry data of \emph{SDSS}(\emph{DR7}) based on the quasar catalogue by \citet{SchneiderETAL2010}.
We randomly selected two subsets of $30,000$ objects each.
One subset was used for training, while the other was used for validation purposes only.
As these experiments are just of conceptual nature, we did not apply cross-folding techniques as we would do, when producing a scientifically relevant catalogue.

\subsection{Effect of Number of Neighbours and Number of Gaussians}

\begin{figure*}
  \includegraphics[width=\textwidth]{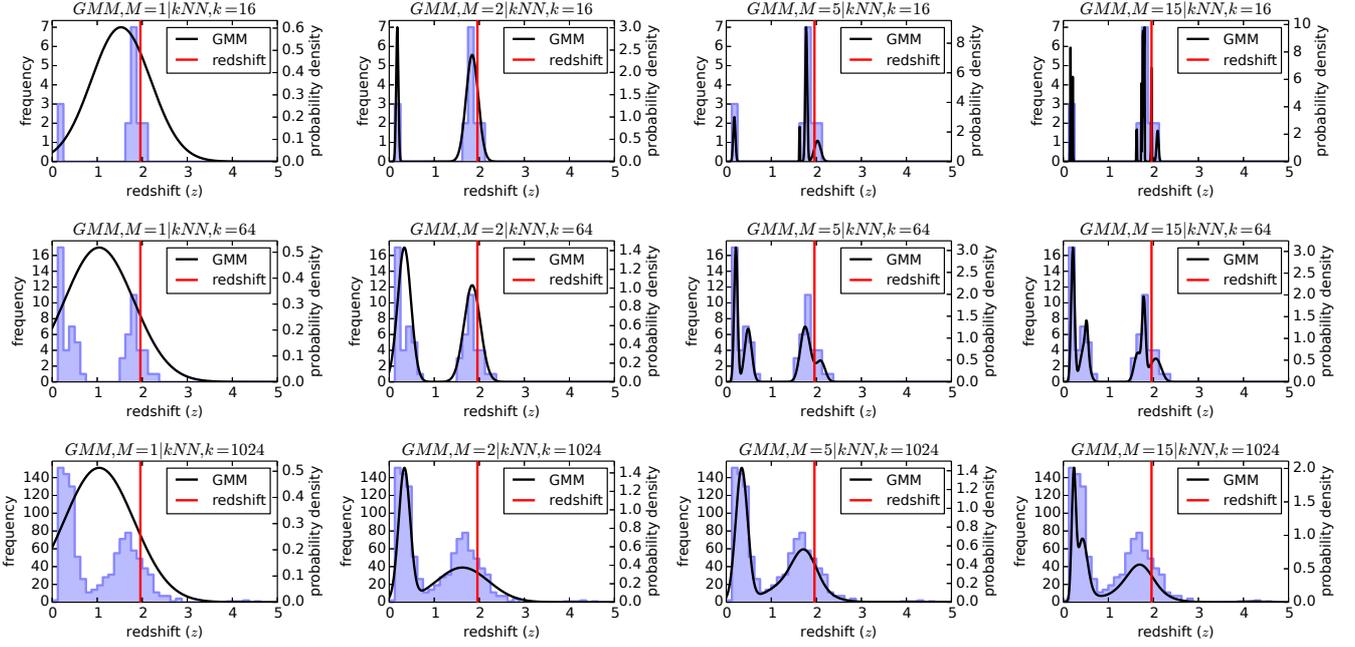}
    \caption{Comparison of resulting \emph{PDF}s optained by varying the number of considered neighbours and varying the number of fitted Gaussian components.
    In this example, 16, 64, and 1024 neighbours are used to fit 1, 2, 5, and 15 Gaussian components, respectively.
    Using just a single Gaussian does not allow to represent the underlying multimodal redshift distribution.
    As one can see, a too low number of considered neighbours with respect to the number of fitted Gaussian components results in a bad representation of the underlying redshift distribution.
    In contrast, too many neighbours smear out the redshift distribution and therefore lead to much broader \emph{PDF}s that exhibit a much lower likelihood at the true redshift (marked red).
    }
    \label{fig:knnComparison}
\end{figure*}

As discussed in the previous section, the number of nearest neighbours is a limiting factor when fitting a \emph{GMM}.
In \mbox{Figure \ref{fig:knnComparison}}, we compare the effects of varying the number of neighbours and the number of Gaussian components.
We used the same object as it is used in \mbox{Figure \ref{fig:nn}}.
With an increasing number of Gaussian components, the resulting \emph{GMM} over-fits.
This is especially true, when the number of available nearest neighbours is too low.
Single redshift values might be represented by a single Gaussian with an extremely low variance.
Therefore the underlying redshift distribution is not represented well and should not be considered for scientific use.
In the multimodal case, a too low number of Gaussian components is generalising too much.
Two separated modes in the redshift distribution cannot be reconstructed by a single Gaussian.
As one can see, the risk of over-fitting the \emph{GMM} decreases with an increasing number of nearest neighbours.
The down side of increasing the number of neighbours is the decrease in sharpness of the redshift distribution.

We found that the best results are optained with a medium number of neighbours and a medium number of Gaussian components.
This number differs as soon as the number of available reference objects and the coverage of the high-dimensional feature space changes.

\subsection{Nearest Neighbour \emph{PDF}s}
\label{nnPDFs}

\begin{figure*}
  \includegraphics[width=\textwidth]{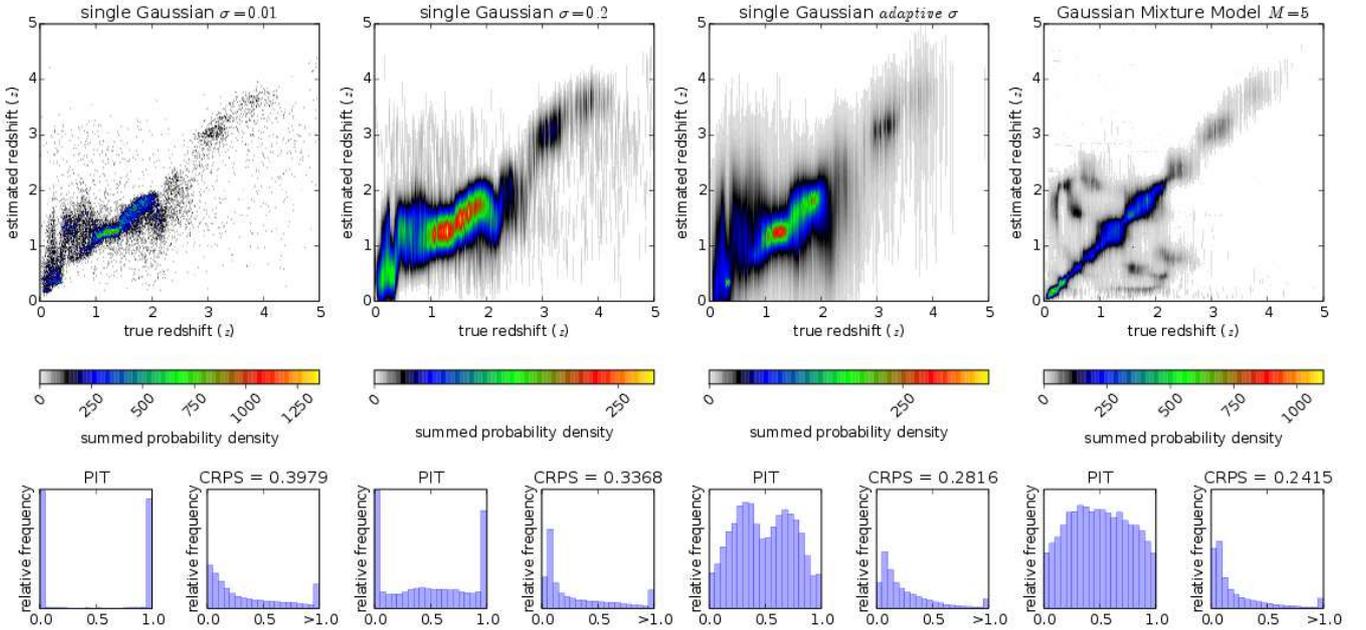}
    \caption{Comparison of different nearest neighbour based \emph{PDF}s.
    The 64 nearest neighbours are used to create (from left to right):
    [i,ii] \emph{PDF}s with a single Gaussian at the mean of the neighbours, with a fixed standard deviation of $\sigma = 0.01$ and $\sigma = 0.2$, respectively.
    [iii] A single Gaussian at the mean of the neighbours with an adaptive standard deviation based on the standard deviation of the neighbours.
    [iv] A \emph{GMM} with five components, fitted to the 64 nearest neighbours.
    Instead of using the usual diagnostic diagram of plotting the true redshift against the estimated one, we plot a density distribution for each object at its true redshift.
    In addition, both the \emph{PIT} and the \emph{CRPS} are presented.
    }
    \label{fig:knnDifferent}
\end{figure*}

We evaluated the \emph{PDF}s that have been generated by four different nearest neighbour based approaches.
In our experiment we choose 64 nearest neighbours to be used to generate the \emph{PDF}s.
In \mbox{Figure \ref{fig:knnDifferent}}, the results of this comparison are presented.
The different approaches are:

\begin{enumerate}
  \item \emph{PDF}s, based on a single Gaussian with the mean of the nearest neighbours and a fixed, very small standard deviation of $\sigma = 0.01$.
  Because of the used extremely narrow single Gaussian, the results can be seen to mimic a single point estimate.
  \item \emph{PDF}s, based on a single Gaussian with the mean of the nearest neighbours and a fixed, broad standard deviation of $\sigma = 0.2$.
  \item \emph{PDF}s, based on a single Gaussian with the mean of the nearest neighbours and an adaptive standard deviation which is the standard deviation of the nearest neighbours.
  \item \emph{PDF}s, based on a \emph{GMM} with five Gaussian components that has been fitted to the nearest neighbours
\end{enumerate}

In order to compare the results, we modified the usually applied diagnostic diagram in which the true redshift is plotted against the estimated one.
Instead of using a single estimated redshift, we use the density distribution of the \emph{PDF}s to plot an intensity parallel to the y-axis.
The resulting plot, represents the average behaviour and the spread of all visualised \emph{PDF}s.
In addition, both, the \emph{PIT} and the \emph{CRPS} are presented in a histogram.

As one can see, the results of the single Gaussians with a fixed, very narrow standard deviation result in a diagnostic plot that is comparable to the ones from point estimates.
The resulting \emph{PIT} shows how extremely underdispersed the resulting \emph{PDF}s are.
The \emph{CRPS} of $0.40$ is the worst of all experiments.

When using the approach with the fixed standard deviation of $\sigma = 0.2$, the \emph{PIT} still exhibits an underdispersion of the \emph{PDF}s.
Even though a quite high standard deviation was chosen, the \emph{CRPS} of $0.34$ indicates a better performance of the \emph{PDF}s, which have been generated with this approach.

The \emph{PIT} histogram of the approach with the adaptive standard deviation shows a slight overdispersion.
This indicates that even with such a large reference set the number of neighbours is not sufficient to sample the redshift distribution.
A \emph{CRPS} value of $0.28$ is a quite big improvement in comparison to the previous experiment.
Due to the underlying physical model, some regions allow to specify the redshift more precisely, e.g. redshift regions where strong spectral features move from one band into another.
Therefore the adaptive specification of the uncertainty is superior to a common uncertainty that is used for all objects.

By allowing for multimodalities, the results improve again.
The \emph{PDF}s that are based on a \emph{GMM} with five components, show both, an improved \emph{PIT} histogram and an improved \emph{CRPS} value of $0.24$.
When inspecting the comparison between the estimated and the true redshifts, the accounting for multimodalities tremendously improves the density concentration towards the ideal diagonal of the plot.
Hence it is important to mention, that such a plot is not a good tool to represent multimodal probability distributions.
We just use this diagnostic plot, as it is very similar to the ones, commonly used for photometric point estimates.

\subsection{Random Forest \emph{PDF}s}

\begin{figure*}
  \includegraphics[width=\textwidth]{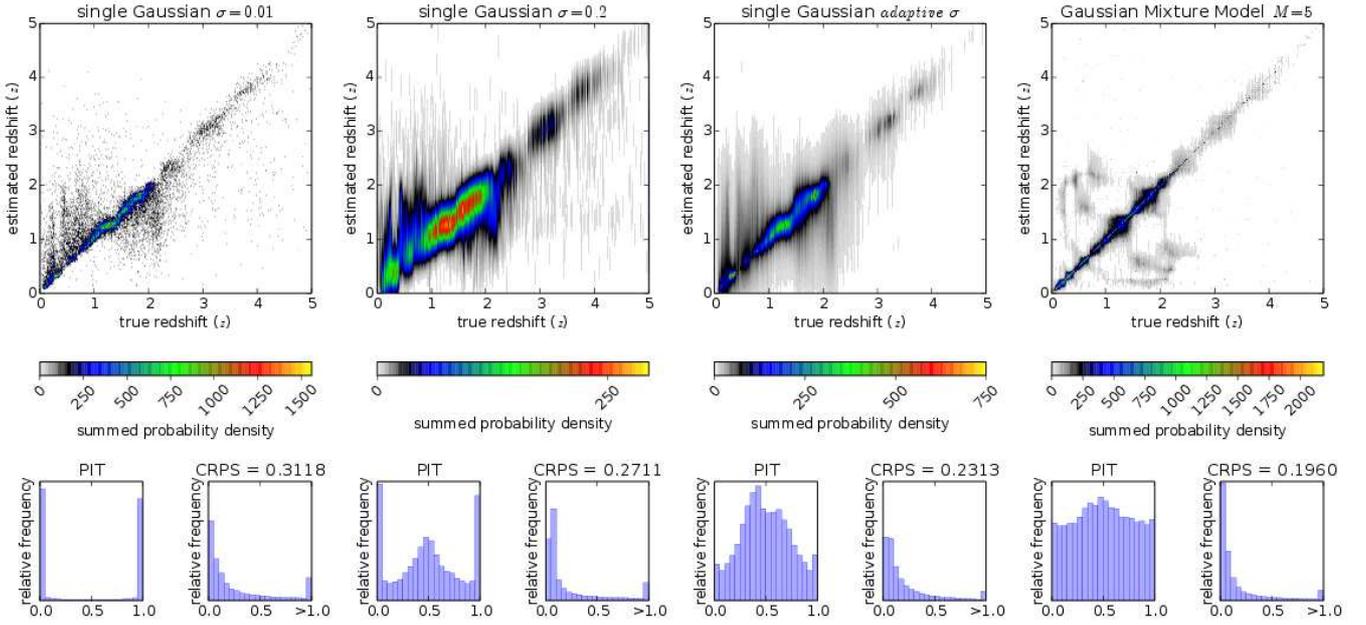}
    \caption{Comparison of different random forest based \emph{PDF}s.
    The results of 256 decision trees are used to create (from left to right):
    [i,ii] \emph{PDF}s with a single Gaussian at the mean of the ensemble, with a fixed standard deviation of $\sigma = 0.01$ and $\sigma = 0.2$, respectively.
    [iii] A single Gaussian at the mean of the ensemble with an adaptive standard deviation based on the standard deviation of the decision trees.
    [iv] A \emph{GMM} with five components, fitted to the results of all trees.
    The same representation as in \mbox{Figure \ref{fig:knnDifferent}} is used.
    }
    \label{fig:rfDifferent}
\end{figure*}

The experiments for measuring the performance of the random forest based \emph{PDF}s was done with an ensemble of 256 decision trees.
We choose four approaches that are similar to the ones used in \mbox{Subsection \ref{nnPDFs}}:

\begin{enumerate}
  \item \emph{PDF}s, based on a single Gaussian with the mean of the ensemble and a fixed, very small standard deviation of $\sigma = 0.01$.
  \item \emph{PDF}s, based on a single Gaussian with the mean of the ensemble and a fixed, broad standard deviation of $\sigma = 0.2$.
  \item \emph{PDF}s, based on a single Gaussian with the mean of the ensemble and an adaptive standard deviation which is the standard deviation of all trees.
  \item \emph{PDF}s, based on a \emph{GMM} with five Gaussian components that has been fitted to the results of the individual trees.
  As discussed above, we could utilize more than 256 trees and therefore fit more Gaussian components, but we want to produce results that are still comparable to the nearest neighbours approach.
\end{enumerate}

In \mbox{Figure \ref{fig:rfDifferent}} the results of these four approaches are summarised.
With the very narrow single Gaussian \emph{PDF}s which mimic point estimates, the random forest approach already produces better results than the comparable nearest neighbours based approach.
This is reflected by a better \emph{CRPS} value of $0.31$.
The reason for this is that already 64 nearest neighbours are producing a mean estimate as they are covering a too large volume in the high-dimensional feature space.
The \emph{PIT} is very efficient in tracing the extreme underdispersion of the \emph{PDF}s produced with the first approach.

The second approach with a fixed but broader standard deviation, produces already a \emph{CRPS} of $0.27$ that is comparable to the adaptive setting in the nearest neighbours experiment.

With an adaptive selection of the standard deviations for the single Gaussian \emph{PDF}s, the multimodal approach of the nearest neighbours is already beaten with respect to the \emph{CRPS}.
For many objects, the nearest neighbours already cover a too large region in the redshift domain and therefore produce much broader \emph{PDF}s than the random forest approach.
The \emph{PIT} histogram of this approach still shows an overdispersion, a clear indicator of a more complex underlying redshift distribution.

When using a \emph{GMM} with five components, we are able to reconstruct the redshifts very well.
The \emph{CRPS} value went down to $0.20$ in this experiment with a \emph{PIT} histogram showing just a slight overdispersion.
In \mbox{Table \ref{tab:results}} the resulting \emph{CRPS} values of all experiments are summarized.

\begin{table}
  \centering
  \caption{Comparison of the \emph{CRPS} values of all experiments.
  }
  \label{tab:results}
  \begin{tabular}{lrr}
    \hline
     & Nearest Neighbours & Random Forest\\
     & (\emph{CRPS}) & (\emph{CRPS})\\
    \hline
    $\mathcal{N}(\mu,\sigma=0.01)$ & $0.3979$ & $0.3118$\\
    $\mathcal{N}(\mu,\sigma=0.2)$ & $0.3368$ & $0.2711$\\
    $\mathcal{N}(\mu,\text{adaptive }\sigma)$ & $0.2816$ & $0.2313$\\
    $GMM, M=5$ & $0.2415$ & $0.1960$\\
    \hline
  \end{tabular}
\end{table}

\section{Conclusion}
\label{conclusion}

The problem with estimating photometric redshifts is the degeneracy of the redshift reconstruction problem, caused by the too low spectral resolution of the photometric features.
We have shown, that in case of photometric redshift estimation, multimodal redshift distribution have to be expected.
A probabilistic description of the redshift estimates allows the scientists to account for this and to propagate the uncertainties correctly.

With the proposed two approaches, \emph{PDF}s can be generated in a straightforward way.
Those \emph{PDF}s can be used to describe photometric redshifts.
As the presented approach is very general in can be applied to any kind of regression problem that should produce probabilistic estimates. 

When dealing with \emph{PDF}s, proper tools for evaluating the performance are required.
We introduced the \emph{PIT} and the \emph{CRPS} as evaluation tools for photometric redshift estimation.
In our experiments we could show, how well those tools help to understand and evaluate probabilistic redshift estimates.
Uncertain photometric redshifts will play an import role very soon in many projects and experiments.
Therefore we would recommend to make the presented tools a standard when comparing \emph{PDF}s in astronomy.

\section*{Acknowledgements}

The authors gratefully acknowledge the support of the Klaus Tschira Foundation.
The authors would like to thank Tilmann Gneiting for the inspiring discussion on how to evaluate the quality of predictive distributions.
The authors appreciate the fruitful discussion with Nikos Gianniotis on the importance of likelihoods.
This work is based on data provided by the SDSS.
SDSS-III is managed by the Astrophysical Research Consortium for the Participating Institutions of the SDSS-III Collaboration.
Funding for SDSS-III has been provided by the Alfred P. Sloan Foundation, the Participating Institutions, the National Science Foundation, and the U.S. Department of Energy Office of Science.
The SDSS-III web site is http://www.sdss3.org/.

%%%%%%%%%%%%%%%%%%%%%%%%%%%%%%%%%%%%%%%%%%%%%%%%%%

%%%%%%%%%%%%%%%%%%%% REFERENCES %%%%%%%%%%%%%%%%%%

% The best way to enter references is to use BibTeX:

\bibliographystyle{mnras}
\bibliography{uncertainRedshift} % if your bibtex file is called example.bib

%%%%%%%%%%%%%%%%%%%%%%%%%%%%%%%%%%%%%%%%%%%%%%%%%%

%%%%%%%%%%%%%%%%% APPENDICES %%%%%%%%%%%%%%%%%%%%%

\appendix

\section{Simple PDF Generation Example}
\label{exampleCode}

In \mbox{Section \ref{generating}} we introduced two simple concepts on how to generate \emph{PDF}s.
Here we give more detailed information on how to implement those concepts in \emph{Python}.

In the first lines of the code we make sure that all required packages are imported.
The {\tt numpy} package provides all matrix and vector manipulation functionalities required for scientific computing.
To solve the regression task, the {\tt scipy.spatial} package and the {\tt sklearn.ensemble.RandomForestRegressor} class are imported.
From the {\tt scipy.spatial} package the k-dimensional tree structure is used to speed up the search for the nearest neighbours.
Finally the {\tt sklearn.mixture.GMM} class is required to express the individual results as a mixture of Gaussians.

\lstset{language=Python}
\begin{lstlisting}[firstnumber=1, title="create\_kNN\_PDFs.py"]
import numpy
import scipy.spatial
from sklearn.ensemble import RandomForestRegressor
from sklearn.mixture import GMM
\end{lstlisting}

We assume that the training data is split into individual files for training and testing the model.
The files are marked with 'X' and 'Y' depending on whether they contain the features or the target values.
All files are loaded from a comma separated value file via a {\tt numpy} function into a matrix or vector, respectively.

\begin{lstlisting}[firstnumber=5]
trainX = numpy.loadtxt("trainX.csv", delimiter=',')
trainY = numpy.loadtxt("trainY.csv", delimiter=',')
testX = numpy.loadtxt("testX.csv", delimiter=',')
testY = numpy.loadtxt("testY.csv", delimiter=',')
\end{lstlisting}

Based on the training data a spacial search structure is initialized.
Next, the search structure is used to determine the redshifts of the nearest 100 neighbours for each of the given test objects.

\begin{lstlisting}[firstnumber=9]
tree = scipy.spatial.KDTree(trainX)
predictions = trainY[tree.query(x=testX, k=100)[1]]
\end{lstlisting}

Finally we fit a Gaussian mixture model with five components to the redshifts of the 100 nearest neighbours.
Note that we limit the minimal covariance to 0.0001 to avoid over-fitting.
This generates an ensemble of five Gaussians with individual weights, means and covariances.
In this example, we just print those values even though they could be used for scientific analysis.
One could easily determine how many components are required for a good fit by making use of the functionalities provided by the {\tt GMM} class.

\begin{lstlisting}[firstnumber=11]
for i in range(len(predictions[:,0])):
  myModel = GMM(n_components=5,
                min_covar=0.0001).fit(predictions[i])
  print myModel.weights_
  print myModel.means_
  print myModel.covars_

\end{lstlisting}

When replacing \mbox{Lines 9-10} by the following lines, a random forest regressor instead of a nearest neighbours approach is used.
First the random forest is created and the 100 decision trees are fitted to represent the given data.
As each individual estimator of the random forest gives an individual result for each input, those results are retrieved individually and stored in a list.
Finally this list of predictions has to be represented as a matrix and transposed to fit the required data-format.

\begin{lstlisting}[firstnumber=9, title="create\_RF\_PDFs.py"]
randomForest = RandomForestRegressor(n_estimators=100)
randomForest.fit(trainX, trainY)

predictions = []
for i in range(len(randomForest.estimators_)):
  predictions.append(
    numpy.array(
      randomForest.estimators_[i].predict(testX)))

predictions = numpy.asarray(predictions).T
\end{lstlisting}

\section{PIT Example Code}
\label{pitExampleCode}

In \mbox{Subsection \ref{PIT}}, the \emph{PIT} was introduced as a powerful tool to analyse the \emph{PDF}s.
This \emph{Python} example shows how to easily create a \emph{PIT} for a given set of predictions.

In the first two lines we import the packages used for scientific computing and for plotting.
Both the {\tt numpy} and the {\tt matplotlib} package are often used in the astronomy community.

\begin{lstlisting}[firstnumber=1, title="create\_PIT.py"]
import numpy
from matplotlib import pyplot
\end{lstlisting}

Next, we create $10,000$ synthetical objects with a random value $y$ between $0.0$ and $6.0$ sampled from an uniform distribution.
As a prediction, we generate $100$ samples from a normal distribution $\mathcal{N}( \mu=0.0,\sigma=0.2)$ for each object, that can be seen as representatives of a \emph{PDF}.
Predictions were shifted with respect to the original values based on deviations that are sampled from a normal distribution with a sigma of 0.2.
In the end we have $10,000$ times $100$ samples from Gaussians that are slightly deviating with their mean from the $y$ values.

\begin{lstlisting}[firstnumber=3]
n = 10000
y = numpy.random.uniform(low=0.0,high=6.0,size=(n))
error = numpy.random.normal(loc=0.0,scale=0.2,size=(n))
predictions = y[:,None] + error[:,None] + numpy.random.normal(scale=0.2,size=(n,100))
\end{lstlisting}

To generate a \emph{PIT} that is based on individual values is as simple as generating it based on a \emph{PDF}.
What is required is the cumulative probability at the value $y$ that is compared with the \emph{PDF}.
To achieve this, the predictions are sorted per object.
Next, the number of predictions below the corresponding $y$ value is determined.
With respect to the number of samples, the cumulative probability at the value $y$ is calculated.

\begin{lstlisting}[firstnumber=7]
predictions = numpy.sort(predictions,axis=1)

PIT = numpy.zeros(n)
for i in range(len(predictions)):
    PIT[i] = len(numpy.where(predictions[i]<y[i])[0]) * 1.0/100.0
\end{lstlisting}

Finally the \emph{PIT} is plotted by simply generating a histogram over all individual cumulative probabilities.

\begin{lstlisting}[firstnumber=12]
pyplot.figure()
pyplot.hist(PIT, bins=10)
pyplot.show()
\end{lstlisting}

%%%%%%%%%%%%%%%%%%%%%%%%%%%%%%%%%%%%%%%%%%%%%%%%%%

% Don't change these lines
\bsp	% typesetting comment
\label{lastpage}
\end{document}